\documentclass[sigconf]{aamas} 

\usepackage{balance} 
\usepackage{pifont} 

\usepackage{amssymb}
\usepackage{newunicodechar}
\usepackage{graphicx} 
\renewcommand\footnotetextcopyrightpermission[1]{}  
\title[AAMAS-2026 Formatting Instructions]{U2F: Encouraging SWE-Agent to Seize Novelty without Losing Feasibility}

\author{Wencheng Ye}
\affiliation{
  \institution{Tongji University}
  \city{Shanghai}
  \country{China}}
\email{2350227@tongji.edu.cn}

\author{Yan Liu}
\affiliation{
  \institution{Tongji University}
  \city{Shanghai}
  \country{China}}
\email{givemeareason@gmail.com}

\begin{abstract}
  Large language models (LLMs) have demonstrated considerable capabilities in software engineering tasks, yet existing LLM-based
  SWE-Agent predominantly address well-defined problems using conventional approaches, potentially overshadowing viable alternative
  solutions that fall outside their predefined frameworks. This limitation becomes notable in open-world software environments where
  novel challenges emerge that extend beyond established solution paradigms. Current SWE-Agents, while making initial attempts at
  addressing such complexities, are constrained by ad-hoc methodologies and tend to operate within narrow scopes—targeting
  localized problems such as individual bug remediation—rather than engaging in comprehensive solution space exploration.

  In this work, we present U2F (Unknown Unknowns to Functional solutions), a cognitive-inspired and uncertainty-embracing multi-agent architecture tailored to systematically surface ``Unknown
  Unknowns''—solution pathways absent from initial problem formulations but possessing innovative potential.
  Our scaffolding comprises two core components. First, we develop a Discovery-Exploration-Integration agent system that enables Unknown Unknowns identification at problem deconstruction,
  creative exploration, and feasible synthesis levels. Second, we implement cognitive enhancement mechanisms along three
  axes—cross-domain analogical reasoning, reverse thinking, and external validation—that observe, analyze, and
  transform conventional solution boundaries through strategic reframing and knowledge integration.

  We demonstrate our framework's effectiveness by applying it to 218 real-world software enabler stories, which were curated from authentic engineering tasks , with the resulting solution quality assessed by 15 domain experts and an LLM judge. Human assessments show improvements: novelty scores
   increased by 14\%, semantic novelty improved by 51\%, while feasibility remained stable at 4.02/5.0, with consistent findings from
  the LLM judge. These results suggest a promising direction for leveraging uncertainty as a source of innovation in software 
  engineering.
  \end{abstract}


\newcommand{\BibTeX}{\rm B\kern-.05em{\sc i\kern-.025em b}\kern-.08em\TeX}

\begin{document}


\pagestyle{fancy}
\fancyhead{}


\maketitle 

\section{Introduction}

While modern software engineering agents powered by large language models
(LLMs) excel at fixing known problems, their ability to innovate—to discover solutions no one thought to ask for—remains underdeveloped. This leaves substantial potential for innovation, grounded in extensive knowledge repositories, largely untapped. Though LLMs possess vast pre-trained knowledge, existing systems face limitations in leveraging this knowledge for creative problem-solving in open-world environments~\cite{anonymous2024,he2024llmbased}, where dynamic, loosely-bounded ecosystems introduce substantial uncertainty and risk~\cite{10.1109/MC.2006.362,OliveiraJr_2024}, requiring solutions that extend beyond conventional approaches.

\begin{figure}[h]
  \centering
  \hspace*{-0.3cm}
  \includegraphics[width=1.05\linewidth]{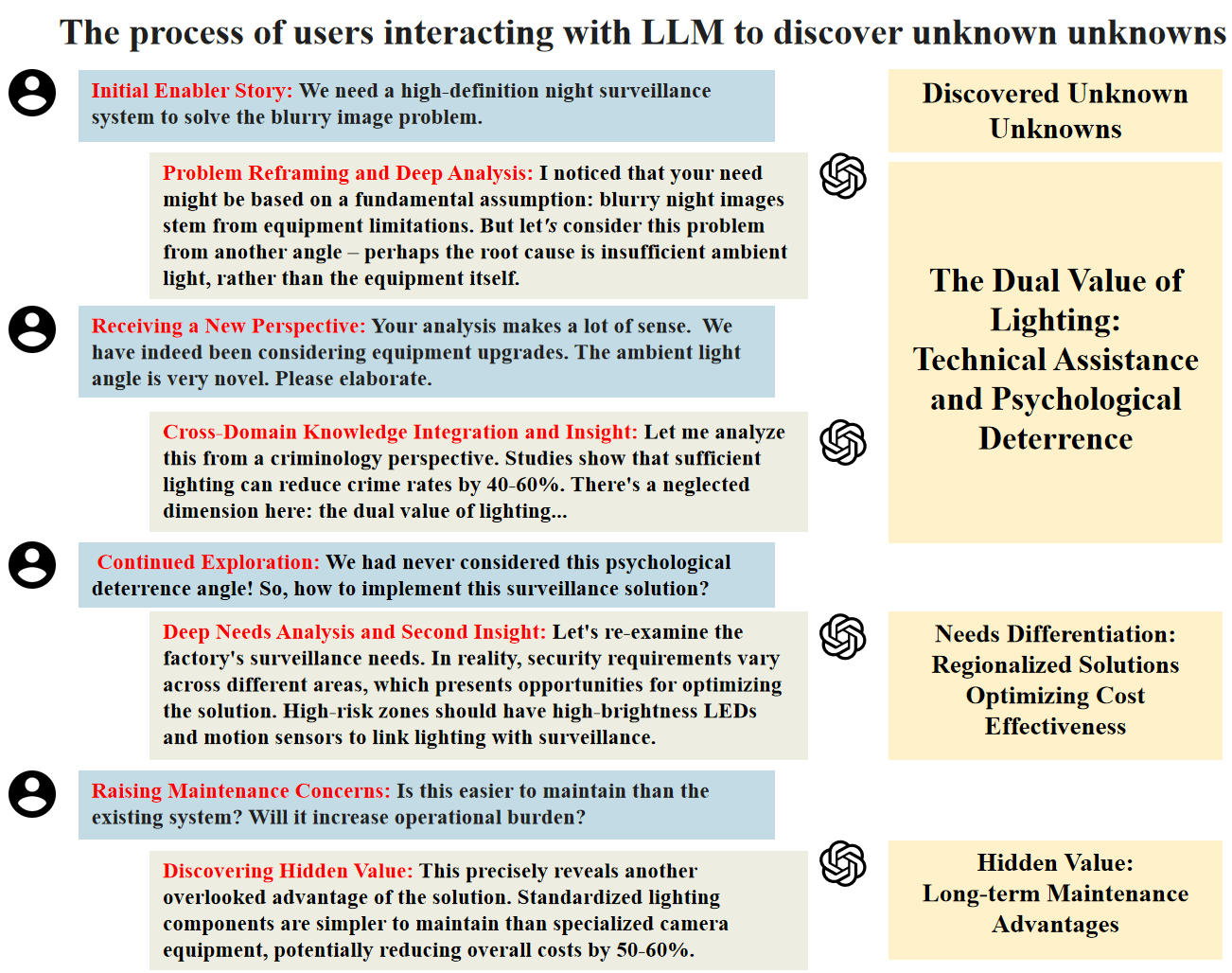}
  \caption{Progressive discovery of Unknown Unknowns. The dialogue demonstrates how systematic exploration reveals solution pathways absent from the initial requirement.}
  \label{fig:llm_example_night_security}
\end{figure}
Classic software engineering practice excels at addressing structured problems within established frameworks—what Rumsfeld's taxonomy categorizes as "Known Knowns" (familiar problems with established solutions) and "Known Unknowns" (recognized challenges requiring investigation) ~\cite{daase2007knowns}. However, this conventional focus may overlook opportunities beyond established problem framings, forming "Unknowns Unknowns" (UUs) ~\cite{bryan2022value, daase2007knowns, mitzen2011knowing}: solution paths, technical approaches, or system capabilities absent from initial problem descriptions but potentially influencing project outcomes. Rumsfeld's framework also identifies "Unknown Knowns"—
knowledge residing in team members' expertise yet unarticulated or unrecognized ~\cite{daase2007knowns}.  Eliciting such implicit knowledge requires probing unarticulated human expertise through interactive methods, a process fundamentally distinct from our framework's automated analysis of documented artifacts.

The potential for such paradigm-shifting insights has also not gone unnoticed by software engineering practitioners, leading to the development of specialized frameworks for technical discovery. While frameworks like SAFe Agile acknowledge this challenge through "Enabler Stories" for exploratory technical work~\cite{scaledagile2023}, their solution design remains heavily dependent on individual engineer expertise, lacking formal approaches for UUs discovery.

Recently, the emergence of LLMs with their vast, interconnected knowledge repositories has sparked
exploration into requirement engineering. 
Consider the development of a night surveillance system, as illustrated in Figure~\ref{fig:llm_example_night_security}. Traditional approaches focus on optimizing detection algorithms or developing sophisticated alarm systems within the established problem frame. However, an Unknowns Unknowns emerges through interacting with LLM: proactive illumination systems can simultaneously serve surveillance needs while providing psychological deterrent effects—a solution pathway absent from the initial technical requirement. This reveals a glimpse of what's possible, yet such 'aha moments' are rare and unpredictable. 

Advances in LLM-based SWE agents further offer promising tools for solution generation~\cite{cheng2025generativeairequirementsengineering,sami2024aibasedmultiagentapproach}. However, directly applying these agents to technical exploration faces critical limitations: technology stacking without clear rationale, requirement drift from original objectives, and generation of non-existent or contextually inappropriate solutions~\cite{jin2024llms, norheim2024challenges}. Addressing these limitations requires more than prompt engineering~\cite{ma2025what}—it demands cognitive-level problem reframing to unlock the innovative potential embedded in LLMs' vast knowledge repositories.

This study explores how LLM  can navigate open-world software environments to surface Unknowns Unknowns through a three-dimensional cognitive approach: \textit{cross-domain analogical reasoning} that transfers insights from diverse fields to reveal unexpected solution pathways; \textit{reverse thinking} that employs backward reasoning from desired outcomes to uncover hidden prerequisites and alternative approaches; and \textit{external validation} that systematically identifies feasibility constraints before they become implementation barriers.

These cognitive pathways lay the foundation of U2F (Unknowns Unknowns to Functional solutions), our framework features a design thinking inspired collaborative architecture: a Discovery Agent that identifies project constraints and exploration opportunities; an Exploration Agent that applies the cognitive strategies above to uncover Unknowns Unknowns; and an Integration Agent that synthesizes discoveries into actionable technical plans while maintaining feasibility. By incorporating external search capabilities and human-in-the-loop interfaces to ground exploration in real-world constraints, we developed a proof-of-concept prototype. 

Through evaluation by 15 domain experts and an LLM judge on 218 carefully curated scenarios, results show that U2F enhances solution innovativeness (14\% improvement in novelty scores, a 51\% increase in semantic novelty) while maintaining engineering feasibility, supporting the viability of systematic Unknowns Unknowns discovery in software engineering contexts.

Our contributions include:

(1) \textbf{Problem Formulation and Initial Exploration:} As far as we know, this is the first work to investigate Unknowns Unknowns discovery in the context of LLM-based SWE-Agents. We establish operational definitions, taxonomies, and exploration methodologies, providing a foundation for future research in this emerging area.

(2) \textbf{Cognitive-Level Innovation Mechanisms:} We introduce and validate three complementary cognitive strategies that operate at the cognitive level rather than merely the prompt level, transforming how LLMs approach open-ended technical requirements.

(3) \textbf{Validation via a Multi-Agent Prototype:} We craft U2F, a rigorous prototype that instantiates these cognitive strategies and agents orchestration. The system demonstrates how coordinated agent architectures can transcend individual cognitive limitations, achieving measurable improvements in solution innovativeness (14\% novelty increase, 1.5× semantic novelty) while maintaining engineering feasibility (4.02/5.0).

\section{Related Work}

\subsection{Unknowns Unknowns in Software Engineering Contexts}

The concept of "Unknowns Unknowns," introduced by Rumsfeld, refers to “things we don’t know we don’t know”~\cite{bryan2022value, daase2007knowns}—the deepest form of uncertainty.  These unknowns involve solution paths or system capabilities that decision-makers may not even recognize~\cite{daase2007knowns, mitzen2011knowing}. Systematically addressing them is crucial in strategic planning, security, and intelligence for improving decisions and reducing systemic risks~\cite{daase2007knowns, mitzen2011knowing}.

Applying this concept to software engineering is useful, as development occurs in complex ecosystems with vast solution spaces, interdependencies, and rapidly evolving technologies. Prior work in requirements engineering has proposed creative elicitation methods, e.g., analogical reasoning, counter-examples~\cite{6636709,sporsem2023knownsunknownsexperiencereport} and building prototypes to tackle unknowns unknowns~\cite{JENSEN20171,KRIESI2016790}, but these remain largely manual, expensive and expertise-dependent. This paper introduces the first structured, LLM-based multi-agent framework for UUs discovery in software engineering, harnessing AI to accelerate large-scale discovery of hidden requirements.

\subsection{Enabler Stories and Double Diamond Model as Vehicles for Innovation Exploration}
Among the many frameworks available for innovation exploration in software engineering, two are particularly influential: Enabler Stories and the Double Diamond model.

In agile software engineering, user stories capture user needs and business value, but often miss implementation complexities and technical exploration. Enabler Stories fill this gap, focusing on architecture, infrastructure, technical debt reduction, and design validation~\cite{dimitrijevic2015comparative}. Unlike user stories, Enabler Stories (“For [target capability], as [role, e.g., development team], we need [technical requirement], so that [technical outcome or next step]”) emphasize technical requirements that enable future capabilities ~\cite{dimitrijevic2015comparative}. SAFe categorizes them as \textit{Architectural}, \textit{Infrastructure}, \textit{Research}, and \textit{Compliance} Enablers. Enabler Stories naturally support Unknown Unknowns exploration by creating systematic opportunities to discover solution paths not specified initially. However, current analyses rely heavily on individual expertise and lack structured frameworks for systematic exploration. Additionally, there is no mature dataset or evaluation mechanism to  support this type of requirement expression.

The Double Diamond model, developed by the UK Design Council, is also a structured framework for fostering innovation through four iterative phases: Discover, Define, Develop, and Deliver ~\citep{article1}. The model is divided into two diamonds: the first (Discover and Define) focuses on exploring  problem space, while the second (Develop and Deliver) concentrates on ideating  solutions~\citep{banfield2016design}. In the \textit{Discover} phase, teams engage in broad research, such as user interviews, observations, and benchmarking, to uncover hidden needs  ~\citep{article1}. The \textit{Define} phase synthesizes these insights to articulate a clear problem statement, reframing assumptions to align with user needs. The \textit{Develop} phase encourages ideation, prototyping, and testing diverse solutions, while the \textit{Deliver} phase refines and implements the most promising ideas, ensuring feasibility ~\citep{saad2020double}.

\subsection{LLM-Based Agents in Requirements Engineering: Capabilities and Limitations}

LLMs enhance requirements engineering across elicitation, analysis, specification, and validation, e.g., extracting requirements from unstructured sources~\cite{arora2024advancing, hemmat2025research}, classifying ambiguities~\cite{marques2024using, lubos2024leveraging}, and automating validation~\cite{arora2024advancing, norheim2024challenges}, by advancing prompting can also elicit implicit requirements~\cite{ma2025what, chen2025promptware}.

 This potential is being realized through the development of SWE agents that extend LLMs beyond
  passive knowledge retrieval to active exploration. Generative agents extend LLMs by autonomously interacting with environments, validating specifications, and generating prototypes~\cite{yang2024sweagent, jin2024llms, cheng2025generativeairequirementsengineering}. However, challenges remain: hallucinations propagate errors~\cite{jin2024llms, bruno2023insights}, requirement drift risks misalignment~\cite{norheim2024challenges}, and uncoordinated technology stacking complicates integration~\cite{jin2024llms}.

While LLMs can occasionally produce "aha moment" insights through spontaneous reasoning~\cite{yang2025understandingahamomentsexternal}, such breakthroughs emerge unpredictably and lack systematic triggers. These moments, though valuable for generating novel insights and leveraging semantic associations across domains, remain difficult to reproduce and their internal mechanisms are opaque. Overcoming this requires novel prompt strategies~\cite{chen2025promptware, ma2025what} and multi-agent architectures~\cite{jin2024llms}, as addressed by our U2F framework. In contrast, our structured UUs discovery approach provides systematic exploration methodologies with quantifiable frameworks, enabling consistent identification of transformative opportunities beyond conventional boundaries.

\section{The U2F Framework: Design, Implementation, and Methodology}

\subsection{Operationalizing "Unknown Unknowns": Definition and Criteria}

 Adopting the definition from previous studies~\cite{6636709,JENSEN20171,KRIESI2016790}, we define UUs as solution-space factors 
systematically discoverable through exploration but absent 
from initial problem formulations—not because they are 
inherently unknowable, but because conventional elicitation 
methods fail to surface them. This concept is a practical proxy rather than complicated philosophy, intended to make this abstract challenge tractable for empirical investigation and tool development.

\textbf{Operational Definition:} At initial scope definition moment \(t_0\), a factor qualifies as a UUs if it satisfies four conditions: (1) \textbf{Evidence absence}—verifiable absence from documented artifacts including product backlogs,architectural decision records, acceptance criteria, risk registers, technical debt lists and user stories; (2) \textbf{Discovery triggering}—emerges through deliberate exploration strategies such as root-cause analysis, prototyping, scenario simulation rather than routine follow-up; (3) \textbf{Solution-space impact}—changes solution architecture, technology choices, or capability priorities (not merely implementation details), often introducing systemic risks like performance bottlenecks, security vulnerabilities, or compliance failures that may lead to project delays, cost overruns, or operational disruptions; (4) \textbf{Non-triviality}—requires conceptual reframing beyond applying standard engineering practices,  arising from stakeholders' subconscious expectations or unforeseen interactions among known system elements. In supplementary material we provide 5 accepted and 3 rejected cases with detailed rationales, demonstrating boundary-case adjudication protocols. These cases aid in converting UUs factors into manageable risks, thereby enhancing project resilience.
\subsection{ The U2F Cognitive Architecture: Design Principles and Collaborative Model}

\begin{figure}[h]
\centering
\hspace*{-0.2cm}
\includegraphics[width=1.05\linewidth]{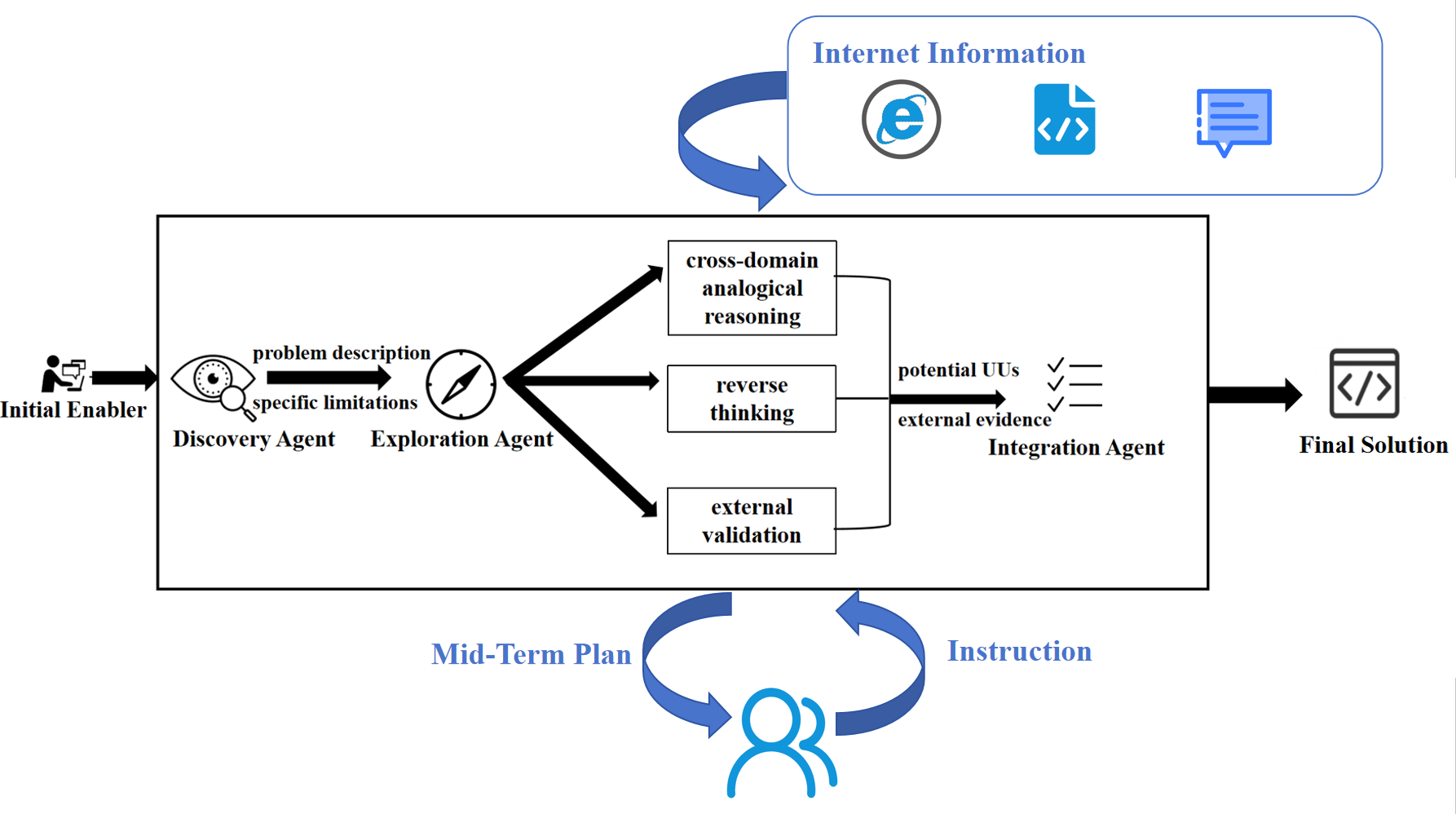}
\caption{Overall architecture of U2F.}
\label{fig:architecture}
\end{figure}

\begin{figure*}[hbt!]
    \centering
    \includegraphics[width=0.85\textwidth]{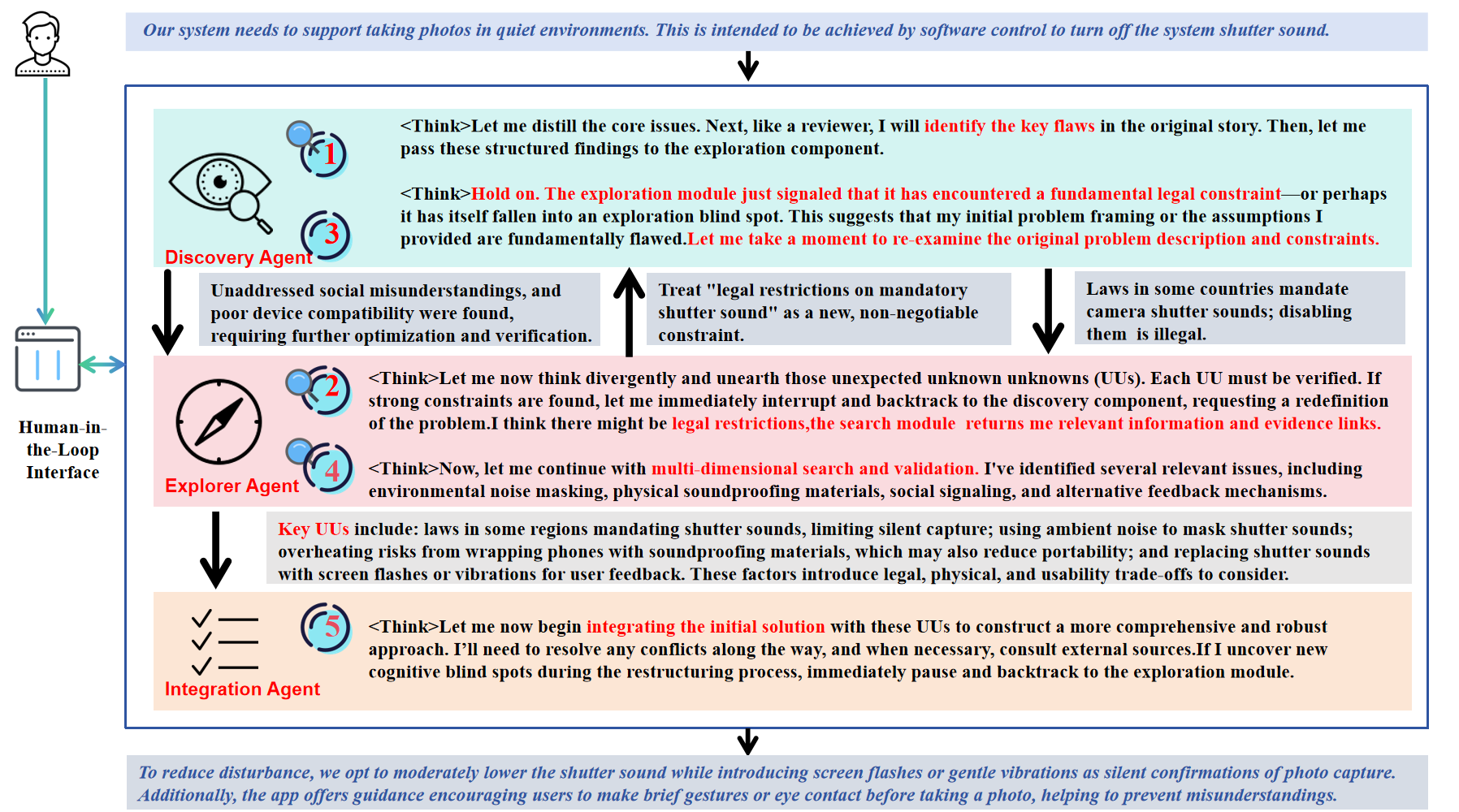}
    \caption{This figure shows a real-world interaction case within our framework, using the context of
developing silent photography in quiet environments. The serial number indicates the processing order of the agent, and the magnifying glass represents the invocation of search API.}
    \label{fig:placeholder}
\end{figure*}
Inspired by the Double Diamond model, U2F leverages 
LLM-based agents to operationalize its phases in software engineering. 
Specifically, as Figure ~\ref{fig:architecture} illustrates, the \textbf{Discovery Agent} mirrors the Discover phase by 
identifying constraints and Unknown Unknowns (UU); the \textbf{Exploration Agent} 
aligns with the {Define} and {Develop} phases by iterating on novel 
ideas; and the \textbf{Integration Agent} embodies the {Deliver} phase by 
ensuring feasible and viable outcomes. This phased orchestration not only enhances 
innovation in software engineering while maintaining practicality, but also 
simulates the cognitive trajectory of human experts when addressing complex and 
ambiguous problems. To realize this, we designed  three core cognitive modules, enabling LLMs to engage in deep, modular 
reasoning across the innovation process.

\subsection{Core Components: From Cognitive Agents to Human-AI Collaboration}
Figure~\ref{fig:placeholder} provides a high-level schematic of this collaborative workflow on a real case. The following subsections will deconstruct this journey, detailing the specific function and design of each component. Full implementation details, including prompt templates and code, are available in the supplementary material.

\subsubsection*{(I) Discovery Component: The Critical Diagnostician}

The Discovery Component acts as the initial diagnostician of the U2F framework. Taking an Enabler Story and its proposed fix as its input, it emulates a seasoned expert's critical eye: first dissecting the problem to its core, then proposing a baseline solution, and finally, rigorously evaluating its inherent weaknesses and blind spots.

Triggered by the user's initial input, this component lays the foundational groundwork for all subsequent innovation. This unfolds in three distinct stages: it first refines the core problem to a concise, actionable statement; then generates a baseline solution from the provided fix; and finally, identifies critical defects by probing for implicit assumptions, scope limitations, and potential side effects. This structured analysis is then handed off to the Exploration Component. However, the Discovery Component remains on standby, ready to be recalled for a strategic reset if later stages reveal a fundamental flaw in the initial problem framing. The deliverable from this stage is a strategic brief: a concise problem description, a summary of the baseline solution, and a clear-eyed analysis of its key limitations and associated risks. 
\subsubsection*{(II) Exploration Component: Mining Potential Key Factors}

Guided by carefully designed prompts, the exploration component constitutes the core innovation engine of our U2F Discover Strategy.  It relies on structured input from the Discovery Component. To formalize this process, we conceptualize exploration as navigation within a three-dimensional open-world cognitive space defined by three orthogonal axes:  the \emph{Analogy–Abstraction} axis, which supports divergent transfer and hypothesis generation across domains; the \emph{Goal–Inversion} axis, which enables backward, goal-directed reasoning to expose minimal prerequisites and hidden dependencies; and the \emph{Epistemic–Validation} axis, which enforces evidence-seeking, contextual grounding, and pragmatic constraints. Accordingly, U2F operationalizes these three cognitive dimensions through the following interrelated strategies:

1. \textbf{Cross-Domain Analogical Reasoning.} This strategy mitigates requirement drift by abstracting the problem into domain-independent representations and searching for analogous structures in unrelated fields, such as biology, psychology, economics, and physics. The analogical mapping can be expressed as:
   $P_{software} \xrightarrow{abstract} P_{general} \xrightarrow{map} S_{domain}$,
   where solutions from diverse domains ($S_{domain}$) are adapted back to the software engineering context. For instance, when examining distributed system reliability, the component might draw inspiration from biological immune system redundancy or economic market stabilization mechanisms. This cross-pollination helps the model escape narrow technical focus, recognize hidden variables, and generate novel yet plausible solution perspectives.

2. \textbf{Reverse Thinking.} To prevent unstrategic technology accumulation, the model performs goal-driven backward reasoning following:
   $G_{target} \leftarrow P_1 \leftarrow P_2 \leftarrow \dots \leftarrow P_{minimal}$,
   where each prerequisite $P_i$ is traced backward from the target goal $G_{target}$ to identify the minimal necessary components. Starting from explicit success criteria, it uncovers dependencies, detects critical bottlenecks, and prunes redundant components. This inversion of the standard forward-chaining process ensures that solutions are anchored in functional objectives rather than technological convenience, while revealing latent risks and implicit assumptions.

3. \textbf{External Validation.}
To minimize hallucinated or ungrounded proposals, each candidate factor is dynamically verified using multiple external evidence sources, including technical literature, implementation case studies, and expert knowledge bases.  
The validation can be represented as 
$
V(f) = F(f) + I(f) + C(f)$, where $F(f)$, $I(f)$, and $C(f)$ denote the assessment of technical feasibility, implementation viability, and contextual appropriateness, respectively.  
All three aspects are considered during validation, grounding creativity in empirical plausibility and ensuring proposals remain practically viable.

The exploration component operates in an adaptive, feedback-driven loop. High-priority or critical Unknowns Unknowns trigger immediate interruption and a call back to the Discovery Component for problem redefinition, effectively forcing the entire strategy to backtrack and restart from a more fundamental analytical level. If multiple factors display potential conflicts, the component defers to the Integration Component for early conflict resolution. Each discovered factor is reported in a standardized three-element format: a concise name and one-line overview, the reason for being overlooked and supporting verification evidence. This structured methodology transforms creative exploration into a rigorous, reproducible framework for identifying Unknowns Unknowns.
\subsubsection*{(III) Integration Component: The Architect of Systematic Redesign}

The Integration Component acts as the architect, weaving the raw "Unknown Unknowns" from the exploration phase into the initial solution. Much like an expert revising their strategy with new insights, this component forges a more robust and innovative final plan. Kicked into action by the Exploration Agent's findings, it synthesizes these new factors while continuously consulting external knowledge sources for the latest technical information. It isn't just a final assembler; it possesses critical self-awareness. If it hits a roadblock—like a new conflict or a flawed initial premise—it can demand deeper exploration or even trigger a full strategic reset by alerting the Discovery Agent.

This integration process unfolds smoothly through four stages. It begins with {Conflict Mapping} to size up how the new UUs challenge or enhance the original plan. Following this, {Solution Refactoring} rebuilds the approach by weaving in new technologies or architectural shifts. Next, {Advantage Attribution} clearly articulates the "why" by pinpointing concrete benefits like cost savings or performance boosts. The process culminates in {Implementation Planning}, which delivers a practical, actionable roadmap complete with technologies, development phases, and risk assessments.

The final deliverable is a concise three-part package: an overview of the new solution's core concepts, a comparative analysis showcasing its advantages, and a detailed implementation plan covering the toolchain, timeline, and potential challenges.

\subsubsection*{(IV) Search Augmentor: External Knowledge Support Throughout the Process}
To overcome the LLM's static knowledge limitations, our framework includes a Search Augmentor that acts as a dynamic, real-time fact-checker. Using the Google Search API, it provides the latest information and verifies domain-specific details. Crucially, this is not a blind trawl for data; it's an on-demand tool activated when an agent explicitly signals a need to supplement facts or validate a hypothesis. The Search Augmentor is woven into the entire process: it helps the Discovery Agent probe the initial solution's weaknesses, enables the Exploration Agent to ground its creative leaps in reality, and provides the Integration Agent with the hard data needed to build a feasible final plan.

\subsubsection*{(V)Human-in-the-Loop Interface: Multi-stage Collaborative Mechanism}
To enhance the system's openness, personalized adaptability, and ensure that the final solution meets the intentions and values of human experts, we introduced a multi-stage human user collaboration mechanism. This interface allows human experts to intervene at key decision points via a command-line interface, guiding and correcting the LLM's reasoning process.

Users can set advanced preferences, taboo conditions, optimization goals (such as "cost first," "innovation first," "minimum risk"), or specify specific thinking paths at any stage. This allows the system to adjust its exploration and generation strategies according to the user's specific business context and strategic goals. Users can also evaluate, question, or fine-tune the output content of each stage through natural language. For example, "This technical solution is too complex, is there a lighter one?", then the system will dynamically adjust subsequent reasoning directions based on this feedback.

This mechanism achieves strategic co-creation and cognitive correction between the intelligent agent and experts, making the entire UU discovery process both model-divergent and rich in expert knowledge, ensuring the innovativeness, feasibility, and high alignment of the final solution with real business scenarios.

\subsection{Building the Experimental Dataset: Generation and Curation of Enabler Stories}
\begin{figure}[H]
\centering
\hspace*{-0.3cm}
\includegraphics[width=1.05\linewidth]{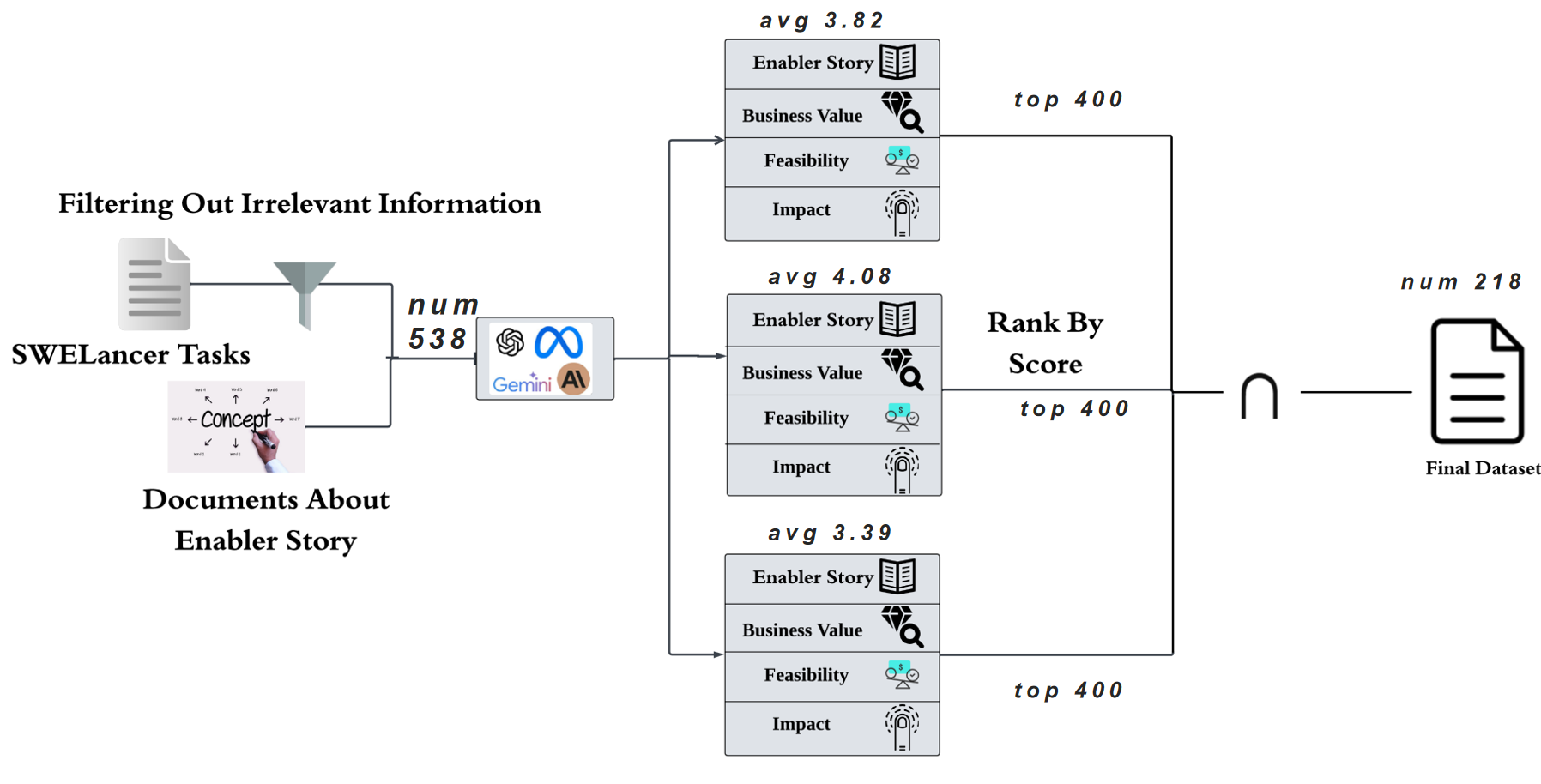}
\caption{Construction process of the Enabler Story dataset.}
\label{fig:data_pipeline}
\end{figure}

To rigorously evaluate the U2F framework's performance on complex, real-world challenges, we first constructed a specialized dataset of Enabler Stories. We adopt an active data construction approach to build an Enabler Story case set. As Figure~\ref{fig:data_pipeline} shows, our process starts with the SWELancer tasks dataset ~\cite{miserendino2025swelancerfrontierllmsearn}, which records authentic but unstructured engineering tasks from real-world developers and testers. To structure these tasks and provide domain context, we inject knowledge from official SAFe documentation ~\cite{scaledagile2023}, offering authoritative definitions and application scenarios for Enabler Stories, guiding the mapping of raw tasks into a conceptual framework.
Structured preprocessing is then applied to extract key causal fields—\textit{Expected Result}, \textit{Actual Result}, and \textit{Potential Fix}—which form microscopic "problem-solution" units suitable for LLM analysis. These fields collectively describe the ideal system behavior, observed deviations or issues, and preliminary solution suggestions. After this preprocessing and screening stage, we retain 538 tasks.
To reduce model-specific bias, we deploy three LLMs (Gemini 2.5-Flash, GPT-4o, Claude 3.7-Sonnet to transcribe the 538 tasks into Enabler Stories. Each task is expanded across four dimensions: a coherent ~200-word narrative describing goal, technical drivers, and story type; \textit{Business Value} (1–5) reflecting potential benefits; \textit{Feasibility} (1–5) considering technical constraints; and \textit{Impact} (1–5) indicating affected modules  dependencies.

Candidate cases that consistently rank in the top 400 across all models are intersected. Crucially, the three-model consensus criterion serves a specific methodological purpose: to identify cases with sufficient technical depth for Unknown Unknowns discovery, rather than trivial code bugs amenable to direct fixes. 
Excluded cases mainly involved simple bug fixes, incomplete requirements, or domain-specific tasks, indicating that the consensus mechanism filtered by problem complexity.
Through this cross-validation screening, we construct a final dataset of 218 cases with high objectivity, diversity, and signal-to-noise ratio, providing a solid foundation for subsequent experiments and analysis.

\section{Experimental Design}

\subsection{Research Questions}

This study aims to answer four main questions:

\begin{itemize}
    \item \textbf{RQ1: Core Method Effectiveness} \\
    Can the U2F framework improve {Novelty} and balance {Feasibility} of solutions compared to initial outputs?

    \item \textbf{RQ2: Comparison with Baselines} \\
    How does our framework perform in terms of {Novelty} and {Feasibility} compared to common prompting strategies: Zero-shot, Role-based, and Software Engineering Analysis Prompting (SEAP)?

    \item \textbf{RQ3: Influencing Factors} \\
    What affects an agent’s ability to generate innovative solutions?
    \begin{itemize}
        \item \textbf{RQ3a (Model Influence):} Differences across foundational LLMs in executing the U2F strategy.
        \item \textbf{RQ3b (Input Type Influence):} Which Enabler Story type (Exploration, Architecture, Infrastructure, Compliance) better inspires innovation?
    \end{itemize}

    \item \textbf{RQ4: Robustness and Reasoning} \\
    How does the framework perform under degraded input quality (e.g., vague, inconsistent requirements)? Can multi-round reasoning improve outcomes?
\end{itemize}

\subsection{Baselines \& Experimental Setup}

  To ensure fully automated and reproducible evaluation, we disable the human-computer interface during experiments and conduct all
  evaluations using stateless API calls to aviod data leakage. Our comparative 
  baselines include:

\begin{itemize}
    \item \textbf{Zero-shot Prompting} \\
    LLM generates solutions directly from Enabler Stories without context or role guidance; serves as minimal benchmark.

    \item \textbf{Role-based Prompting} \\
    LLM assumes a professional role (e.g., senior architect) to guide perspective-driven solution generation.

    \item \textbf{ SEAP(Software Engineering Analysis
    Prompting)} \\
    LLM receives structured, multi-step guidance combining requirements analysis, functional decomposition, risk-driven development, domain-driven design, infrastructure-as-code, and formal verification to  produce feasible, innovative solutions. This baseline represents best practices in systematic software engineering and serves as a strong comparison point.
\end{itemize}

\subsection{Evaluation Metrics}
We adopt a multidimensional evaluation that integrates human judgment and LLM-based blind assessment. A balanced panel of 15, with an equal representation of software experts and students, alongside Qwen-3-32B~\cite{yang2025qwen3technicalreport} as an LLM evaluator, rated all solutions on a 1–5 Likert scale. The panel's composition—featuring seasoned experts (all possessing at least five years of software engineering experience) and undergraduate students majoring in software—ensured a comprehensive evaluation that captured engineering feasibility alongside a fresh, innovative perspective. Prior to the formal evaluation, all panel members participated in a calibration session where they reviewed sample cases and aligned their understanding of novelty and feasibility dimensions, ensuring consistency in evaluation standards. For the LLM evaluator, we employed Qwen-3-32B, different from the models used during dataset construction, to provide independent assessment and reduce potential evaluation bias. We also interviewed the panel members to gather their opinions on this agent-enabled approach.
\begin{table*}[t]
\centering
\caption{Comprehensive Performance Comparison Across Methods (Randomized LLM Scores)}
\label{tab:1}
\resizebox{\textwidth}{!}{
\begin{tabular}{lccccccc}
\toprule
Method & Novelty (Human / LLM) & Feasibility (Human / LLM) & Semantic Novelty & UUs Discovered & Expert Approval & LLM Approval \\
\midrule
Zero-shot Prompting & 2.85 $\pm$ 0.72 / 3.05 $\pm$ 0.66 & 3.55 $\pm$ 0.68 / 3.61 $\pm$ 0.65 & 0.28 $\pm$ 0.16 & 0.25 $\pm$ 0.50 & 0.22 & 0.33 \\
Role-based Prompting & 3.20 $\pm$ 0.69 / 3.51 $\pm$ 0.62 & 3.70 $\pm$ 0.65 / 3.81 $\pm$ 0.60 & 0.35 $\pm$ 0.17 & 0.48 $\pm$ 0.75 & 0.35 & 0.46 \\
SE Analysis Prompting & 3.55 $\pm$ 0.65 / 3.76 $\pm$ 0.64 & 3.90 $\pm$ 0.55 / 4.03 $\pm$ 0.58 & 0.45 $\pm$ 0.19 & 0.80 $\pm$ 0.98 & 0.48 & 0.59 \\
\textbf{U2F} & \textbf{4.05 $\pm$ 0.75 / 4.23 $\pm$ 0.68} & \textbf{4.02 $\pm$ 0.63 / 4.19 $\pm$ 0.61} & \textbf{0.68 $\pm$ 0.18} & \textbf{1.55 $\pm$ 1.05} & \textbf{0.62} & \textbf{0.75} \\
\bottomrule
\end{tabular}
}
\end{table*}

To assess \textbf{Novelty}, we complement expert and LLM ratings with a semantic metric. Expert and LLM evaluations measure methodological, technological, and process novelty on a 1--5 scale, capturing conceptual depth and innovative potential. The semantic metric is defined as
\begin{equation*}
N_{\text{semantic}}(S_{\text{result}}) = 1 - \text{\tiny COSINE\_SIMILARITY}(\text{embed}(S_{\text{result}}), \text{embed}(S_{\text{initial}})),
\end{equation*}
where $\text{embed}(\cdot)$ denotes a Sentence-Transformer model(all-MPNet-base-v2) embedding. Higher values indicate greater innovation relative to the initial solutions, providing quantitative evidence of lexical divergence from baseline solutions. Overall, expert and LLM assessments serve as primary validity criteria, while semantic novelty offers complementary quantitative support.

\textbf{Feasibility} is rated independently by experts and the LLM evaluator, both considering technical implementability, resource rationality, risk control, and compatibility with existing systems and business processes, on a 1--5 scale.

\textbf{Assessing UU Discovery.} We evaluate the framework's core discovery performance using two complementary metrics. First, to measure \emph{quantity}, we count the average number of distinct UUs documented per case (shown as \textbf{UUs Discovered} in Table 1). Second, to assess the \emph{quality} and \emph{relevance} of these findings, each discovered UU is independently judged by human experts and the LLM evaluator. The \textbf{Approval Rate} (shown as \textbf{Expert/LLM Approval}) represents the percentage of discovered UUs that were deemed valid and insightful by each group, thus acting as a quality filter on the raw number of discoveries.

\textbf{Robustness.} We assess the framework’s ability to perform under adverse conditions by evaluating it on a dedicated test set of 50 low-quality samples, designed to simulate real-world data imperfections with 25\% to 60\% of information intentionally removed or obscured. The evaluation of the resulting outputs focuses on three criteria: the \textbf{Failure Rate}, a quantitative measure of the percentage of outputs rendered unusable by logical or structural defects; the \textbf{Logical Coherence}, an expert-rated score (1--5) for internal consistency and clarity; and finally, the solution's \textbf{Relevance} to the original problem's core goals (also rated 1--5), which verifies that the framework was not sidetracked by the noisy input.

\section{Results}
\label{sec:results}

\subsection{Overall Performance Analysis}
\label{subsec:overall-performance}

The results in Table \ref{tab:1} demonstrate U2F's consistent superiority across all evaluation dimensions. While it achieves a 14\% improvement in novelty over SEAP, the semantic novelty metric reveals a substantial shift from 0.45 to 0.68. This suggests U2F guides models toward conceptual reframing, not merely incremental refinement. Human–LLM agreement was strong, with Pearson correlations of 0.81 for novelty and 0.78 for feasibility, alongside Spearman correlations of 0.79 and 0.76, indicating reliable and aligned evaluations. Inter-rater reliability among the 15 experts also showed substantial agreement with Fleiss' Kappa at 0.65.

A key finding is the framework's balance between innovation and feasibility. Despite pursuing high-risk unknown unknowns, it achieves the highest feasibility score of 4.02. This reflects a cognitive process where an exploration phase ventures into uncertain domains, while an integration phase anchors radical ideas in reality. U2F identifies an average of 1.55 potential UU per case, nearly doubling the next-best approach. 

The 62\% expert approval rate demonstrates the 
"explore broadly, filter collaboratively" design philosophy 
aligns with design thinking principles: frameworks proposing 
only safe solutions cannot discover genuine unknowns. Though analyzing, we discovered rejected UUs primarily stem from three factors: 
over-abstraction addressable through domain-specific 
constraint tuning, validation thresholds adjustable via 
search augmentation parameters, and context sensitivity 
improvable through enhanced project metadata. These represent 
engineering refinement opportunities rather than fundamental 
limitations. In real-world deployment, the approval rate suggests that for every 10 Enabler Stories processed, teams can expect approximately 15 UU candidates, of which around 9 merit serious consideration, expand the solution space.

Another important observation from comparing human and LLM scores is that the LLM evaluator consistently rates slightly higher,  yet maintains stable relative ranking across methods. This validates its use for comparative evaluation while necessitating human oversight for absolute scoring. 

\subsection{Research Questions Analysis}

\subsubsection{RQ1: Transformation from Known to Unknown}
\label{subsubsec:rq1}

The comparison between U2F and zero-shot solutions reveals a significant cognitive transformation. Novelty improves by 42.1\%, from 2.85 to 4.05, while semantic novelty surges by 143\%, from 0.28 to 0.68, providing quantitative evidence of its ability to transcend the model's default reliance on familiar patterns. The LLM evaluator mirrors this trend with consistent score elevation from 3.05 to 4.23, reaffirming that the observed performance gains reflect genuine cognitive enhancement rather than evaluator-specific artifacts. Initial zero-shot solutions, while feasible, suffer from "convergent poverty"—they rapidly settle on obvious approaches. U2F's architecture is designed to deliberately disrupt this premature convergence before coalescing possibilities into actionable solutions.

\subsubsection{RQ2: Cognitive Scaffolding Architecture}
\label{subsubsec:rq2}

The progressive performance improvements across methods illuminate the mechanics of cognitive scaffolding. Zero-shot prompting achieves a novelty score of 2.85, capturing the model's unstructured baseline capability. Role-based prompting reaches 3.20, introducing a first layer of structure through perspective assignment. SE Analysis Prompting attains 3.55, representing traditional systematic thinking that excels at methodical problem decomposition but operates within established analytical paradigms. U2F achieves 4.05, transcending this with meta-cognitive awareness that explicitly challenges the problem formulation itself. LLM evaluator's progressive score increases from 3.05 to 3.51 to 3.76 and finally to 4.23 closely parallel human judgments, with the optimistic bias remaining proportionally consistent. This trend indicates that the innovativeness of the output consistently improves as the guiding framework evolves from unstructured prompting to a systematic architecture, which validates that the framework's cognitive scaffolding effectively guides LLM reasoning toward genuinely innovative solutions. The critical distinction is its relationship to uncertainty: while SE Analysis seeks to minimize it, U2F embraces it as a source of innovation.
\setcounter{secnumdepth}{4} 
\setcounter{tocdepth}{4}    

\subsubsection{RQ3: Architectural Robustness and Contextual Sensitivity}

In Cross-Model Generalizability Analysis, U2F performance is remarkably consistent across different foundation models, as shown in Table \ref{tab:model-comparison}. 

This suggests its cognitive scaffolding operates at a level of abstraction that transcends model-specific capabilities. GPT-4o's slight edge in novelty may reflect its training emphasis, while Gemini 2.5 offers a 3.6-fold cost advantage at \$0.05 versus \$0.18 per case without major performance sacrifice, making it attractive for high-volume deployments. Across all three models, the LLM evaluator maintains a consistent positive bias of 0.10 to 0.15 points for novelty and approximately 0.13 for feasibility, further confirming that the scoring elevation is evaluator-intrinsic rather than method-dependent or model-dependent.

\begin{table}[h]
\small
\caption{Performance Consistency Across Foundation Models}
\centering
\label{tab:model-comparison}
\resizebox{\columnwidth}{!}{%
\begin{tabular}{lccccc}
\toprule
Model & Novelty (Human / LLM) & Feasibility (Human / LLM) & UUs Discovery & Cost/Case \\
\midrule
GPT-4o & $4.15 \pm 0.70 / 4.30 \pm 0.65$ & $4.05 \pm 0.60 / 4.18 \pm 0.58$ & $1.62 \pm 1.00$ & \$0.18 \\
Claude 3.7 & $4.08 \pm 0.72 / 4.20 \pm 0.68$ & $3.97 \pm 0.65 / 4.10 \pm 0.62$ & $1.50 \pm 1.08$ & \$0.23 \\
Gemini 2.5 & $3.95 \pm 0.78 / 4.05 \pm 0.70$ & $4.15 \pm 0.68 / 4.25 \pm 0.65$ & $1.45 \pm 1.02$ & \$0.05 \\
\bottomrule
\end{tabular}%
}
\end{table}
In Task Context Sensitivity Analysis, innovation potential is linked to problem structure, as demonstrated in Table \ref{tab:story-types}. "Exploration" tasks, with ambiguous requirements, provide the most fertile ground for UU discovery. Conversely, "Compliance" tasks, constrained by regulations, naturally limit the scope for radical reimagining. However, U2F still achieves meaningful improvement in these rigid domains, suggesting innovation is possible with more sophisticated discovery techniques.

\begin{table}[h]
\centering
\footnotesize
\setlength{\tabcolsep}{4pt} 
\caption{Innovation Potential Across Enabler Story Types}
\label{tab:story-types}
\resizebox{0.3\textwidth}{!}{
\begin{tabular}{lcc}
\toprule
\textbf{Task Type} & \textbf{Avg. Novelty} & \textbf{UUs Discovered} \\
\midrule
Exploration    & $4.25 \pm 0.65$ & $1.90 \pm 1.15$ \\
Architecture   & $3.95 \pm 0.70$ & $1.40 \pm 1.00$ \\
Infrastructure & $3.80 \pm 0.72$ & $1.25 \pm 0.90$ \\
Compliance     & $3.60 \pm 0.80$ & $1.05 \pm 0.85$ \\
\bottomrule
\end{tabular}%
}
\end{table}

\subsubsection{RQ4: System Resilience and Failure Modes}

The robustness analysis exposes the framework's resilience and its limits, as shown in Table \ref{tab:robustness}. Under mild adversarial conditions, U2F demonstrates graceful degradation, maintaining 85\% of its novelty. This stems from its multi-stage design, where later stages can compensate for degraded inputs. However, performance collapses non-linearly beyond a moderate adversarial intensity, with failure rates jumping from 12\% to 25\%. This reveals that when input signal-to-noise ratios are too low, the cognitive scaffolding can become counterproductive. This underscores that sophisticated reasoning requires correspondingly coherent inputs.

\begin{table}[H]
\centering
\caption{Performance Under Adversarial Conditions}
\label{tab:robustness}
\resizebox{\columnwidth}{!}{ 
\begin{tabular}{lcccc}
\toprule
\textbf{Input Quality} & \textbf{Failure Rate} & \textbf{Logical Coherence} & \textbf{Relevance Score} & \textbf{Novelty Retention} \\
\midrule
Standard Conditions & 5.5\% & $4.05 \pm 0.70$ & $4.10 \pm 0.65$ & 100\% \\
Mild Degradation & 12.0\% & $3.70 \pm 0.78$ & $3.80 \pm 0.72$ & 85.0\% \\
Moderate Degradation & 25.0\% & $3.10 \pm 0.85$ & $3.25 \pm 0.80$ & 68.0\% \\
Severe Degradation & 45.0\% & $2.50 \pm 0.90$ & $2.65 \pm 0.88$ & 48.0\% \\
\bottomrule
\end{tabular}
}
\end{table}

\subsection{Ablation Study}
\label{subsec:ablation-study}
\begin{table}[H]
\centering
\caption{Component Contribution Analysis}
\label{tab:ablation}
\resizebox{\columnwidth}{!}{ 
\begin{tabular}{lcccc}
\toprule
\textbf{Configuration} & \textbf{Novelty} & \textbf{Feasibility} & \textbf{UUs Discovery} & \textbf{Performance Impact} \\
\midrule
Complete U2F & 4.05 & 4.02 & 1.55 & Baseline \\
w/o Search Augmentation & 3.60 & 3.90 & 1.20 & -11.3\% novelty \\
w/o Exploration Stage & 3.30 & 4.00 & 0.75 & -18.5\% novelty \\
w/o Integration Stage & 3.75 & 3.80 & 1.30 & -7.4\% novelty \\
Discovery Stage Only & 3.00 & 3.70 & 0.35 & -25.9\% novelty \\
\bottomrule
\end{tabular}
}
\end{table}

The ablation study in Table \ref{tab:ablation} illuminates the distinct functional roles within the architecture. The Exploration stage is the primary innovation engine; its removal triggers the steepest novelty decline of 18.5\% and halves UU discovery. Its use of cross-domain analogies and assumption-challenging is crucial for breaking conventional patterns. The Integration stage acts as a crucial reality bridge; its absence impacts feasibility significantly, dropping from 4.02 to 3.80, and prevents the framework from generating impractical ideas. Search Augmentation provides the factual foundation for both innovation and feasibility. The poor performance of the Discovery-only configuration, with a 25.9\% novelty decline, validates that innovation requires both divergent exploration and convergent synthesis.

\subsection{Practical Impact and Implementation Insights}
\label{subsec:implications}

U2F reduces the otherwise time-intensive task of UU discovery to about {2 to 3 minutes} per case. The framework incurs API costs between \$0.05 and \$0.23 per case depending on model choice, which is considerably lower than the labor costs of senior experts, typically estimated at \$100 to \$300 per story for 1 to 2 hours of review time. It should be noted that the framework's results  require extra cost of expert verification, still
 U2F  substantially alleviate the burden of early-stage investigation and exploratory analysis, enabling experts to focus on higher-value tasks. Over 80\% of experts recognized its role in offering new perspectives and breaking thinking inertia. We invited experts to experience the human-in-the-loop interface,
 73\% reported noticeable improvement, we consider it a critical feature for the framework's practical deployment to its full potential.
 
\section{Conclusion}
In this study, we proposed a multi-component agent system for
systematically discovering "Unknown Unknowns" in software engineering. Future research should enhance UU validation precision
through more sophisticated external knowledge integration and
focus on quantifying the specific value added by human-expert
collaboration at different stages and exploring the optimal cognitive workflow for such a human-AI partnership, developing ethical
guidelines for responsible deployment in production environments.
Overall, our framework provides a foundation for future research
at the intersection of AI and innovative software engineering, suggesting that profit not only from eliminating uncertainty, but also
from exploiting it as a springboard for discovering possibilities.




\bibliographystyle{ACM-Reference-Format} 
\bibliography{sample}

\end{document}